\documentclass{article}

\usepackage{arxiv}

\usepackage[utf8]{inputenc} 
\usepackage[T1]{fontenc}    
\usepackage[colorlinks=true,allcolors=blue]{hyperref}
\usepackage{url}            
\usepackage{booktabs}       
\usepackage{amsfonts}       
\usepackage{amsmath}
\usepackage{mathtools}
\DeclareMathOperator*{\argmin}{arg\,min}
\usepackage{nicefrac}       
\usepackage{microtype}      
\usepackage{lipsum}		
\usepackage{graphicx}
\usepackage[numbers]{natbib}
\usepackage{doi}

\title{Topological potentials guiding protein self-assembly}

\author{ \href{https://orcid.org/0009-0007-0417-4058}{\includegraphics[scale=0.06]{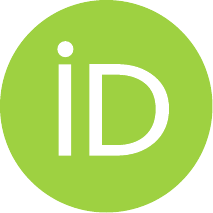}\hspace{1mm}Ivan Spirandelli} \\
	Institute of Mathematics\\
	University of Potsdam\\
	Karl-Liebknecht-Str. 24-25, 14476 Potsdam\\
	\texttt{spirandelli@uni-potsdam.de} \\
    \And
	\href{https://orcid.org/0000-0003-4823-5311}{\includegraphics[scale=0.06]{orcid.pdf}\hspace{1mm}Arnur Nigmetov} \\
	Lawrence Berkeley National Laboratory\\
	1 Cyclotron Rd, Berkeley, CA 94720 \\
	\texttt{anigmetov@lbl.gov} \\
    \And
	\href{https://orcid.org/0000-0002-4330-6670}{\includegraphics[scale=0.06]{orcid.pdf}\hspace{1mm}Dmitriy Morozov} \\
	Lawrence Berkeley National Laboratory\\
	1 Cyclotron Rd, Berkeley, CA 94720 \\
	\texttt{dmitriy@mrzv.org} \\
    \And
	\href{https://orcid.org/0000-0002-0161-6523}{\includegraphics[scale=0.06]{orcid.pdf}\hspace{1mm}Myfanwy E. Evans} \\
    Institute of Mathematics\\
	University of Potsdam\\
	Karl-Liebknecht-Str. 24-25, 14476 Potsdam\\
	\texttt{evans@uni-potsdam.de} \\
}

\renewcommand{\headeright}{}
\renewcommand{\undertitle}{}
\renewcommand{\shorttitle}{Topological potentials guiding protein self-assembly}

\hypersetup{
pdftitle={Topological potentials guiding protein self-assembly},
pdfsubject={q-bio.NC, q-bio.QM},
pdfauthor={Ivan Spirandelli, Dmitriy Morozov, Arnur Nigmetov, Myfanwy E. Evans},
pdfkeywords={self-assembly, kinetic trapping, persistent homology, topological data analysis, protein assembly},
}

\begin{document}
\maketitle

\begin{abstract}
The simulated self-assembly of molecular building blocks into functional complexes is a key area of study in computational biology and materials science.
Self-assembly simulations of proteins using physically-motivated potentials for non-polar interactions, 
can identify the biologically correct assembly as the energy-minimizing state.
Short-range potentials, however, produce rugged energy landscapes, which lead to simulations becoming trapped in non-functional local minimizers.

Successful self-assembly simulations depend on the physical realism of the driving potentials as well as their ability to efficiently 
explore the configuration space.

We introduce a long-range topological potential, quantified via weighted total persistence, and combine it with the morphometric approach to solvation-free energy.

This combination improves the assembly success rate in 
simulations of the tobacco mosaic virus dimer and other protein complexes by up to sixteen-fold compared with the morphometric model alone. 
It further enables successful simulation in systems that don't otherwise assemble during the examined timescales.

Compared to previous topology-based work, which has been primarily descriptive, our approach uses topological measures as an active 
energetic bias that is independent of electrostatics or chemical specificity and depends only on atomic coordinates. 
Therefore, the method can, in principle, be applied to arbitrary systems where such coordinates are optimized. 
Integrating topological descriptions into an energy function offers a general strategy for overcoming kinetic barriers in molecular simulations, 
with potential applications in drug design, materials development, and the study of complex self-assembly processes.
\end{abstract}


\section{Introduction}
In silico assembly of molecular building blocks into functional complexes is a central challenge in biology and materials science. A primary obstacle is the problem of kinetic trapping \cite{kinetic_traps, kinetic_traps_2}: even when the functional native structure represents the minimum of the energy function that governs the simulation, the vastness of the configurational landscape, containing many non-functional local minima, can prevent simulations from discovering it on a feasible timescale. A successful simulation methodology must therefore not only be based on an accurate energy function that identifies the native state, but also navigate it's landscape efficiently to avoid dead-end assembly pathways.

Here we introduce a long-range potential based on the principles of computational topology to act as a guiding bias. Topological data analysis, particularly persistent homology \cite{pers_hom_1, pers_hom_2}, is utilized to solve optimization tasks in machine learning contexts \cite{ph_opt,big_steps}, and has been widely used to analyze static and dynamic molecular data \cite{binding_1, binding_2, structural_1, ph_bio_1} and nanomaterials \cite{ph_mat_sci_1, ph_mat_sci_2} post hoc. We use it here as an active component within the simulation of protein assembly. 

The physically motivated core of our simulations is the \textit{morphometric approach to solvation-free energy} \cite{morpho_1,morpho_2}. It is an implicit solvent model \cite{implicit_solvent_0, implicit_solvent_1} and a sophisticated implementation, describing non-polar contributions to the solvation-free energy as a linear combination of four geometric measures: the excluded volume $V$, solvent accessible surface area $A$, integrated mean curvature $C$, and integrated Gaussian curvature $X$ of the solvent accessible surface \cite{sasa_1, sasa_2, sasa_3}. This method has proven highly accurate for calculating solvation properties of complex biomolecules \cite{morpho_3, morpho_4, morpho_5, evans2014b} and assembles exotic geometries for generic solutes \cite{spirandelli_2024, coles2024solventstieknotshelical}. For simulation of macromolecules, we augment it with a soft-sphere overlap penalty $L$, yielding a total short-range geometric potential:
\begin{equation}
\label{fsol_star}
  F^*_{sol} = p V + \sigma A + \kappa C + \overline{\kappa} X + \phi L.
\end{equation} 
The prefactors for the geometric terms are derived from fundamental measure theory \cite{white_bear}, where $p$ is pressure, $\sigma$ is surface tension, and $\kappa$ and $\overline{\kappa}$ are bending rigidities \cite{morpho_2}. The weight $\phi$ is empirically determined and, in combination with $L$, acts both as a simple model of repulsion and as a relaxation of the shape of the modeled molecules.

Previous work has shown that for tobacco mosaic virus (TMV) subunits, the assembly observed experimentally corresponds to the minimum of $F_{sol}^*$ obtained via Random Walk Metropolis simulations \cite{spirandelli_2025}. This indicates the thermodynamic accuracy of the method. However, this does not guarantee kinetic accessibility. The morphometric approach is inherently short-ranged, as its geometric measures are defined by a local solvent probe (e.g., $r_s = 1.4\text{\AA}$ for water). Consequently, its energy landscape is largely flat until the subunits are in close contact, at which point it becomes rugged. A simulation driven only by $F_{sol}^*$ performs a largely random search until protein subunits interact at a close range. Here, the simulations find local minimizers that are sometimes difficult to escape from, preventing further exploration and correct assembly. This highlights a kinetic, rather than thermodynamic, hurdle that the model has to overcome in a simulation context.

\begin{figure}[ht!]
\centering
\includegraphics[width=8.7cm]{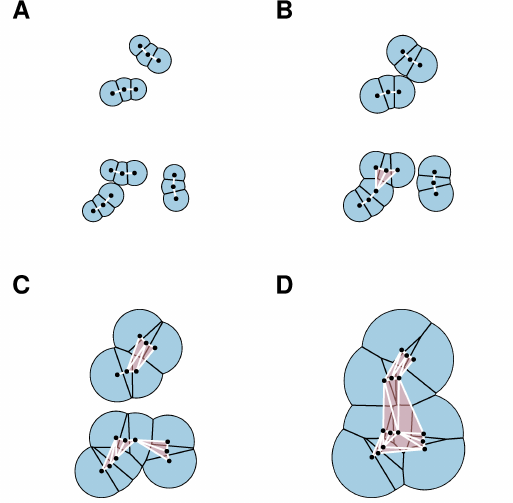}
\caption{Weighted Alpha complexes for system of five toy shapes in two dimensions and four different $\alpha$-values (A-D). Vertices are shown in black, edges in white and triangles in red. The dual intersection of balls with the weighted Voronoi-diagram is shown in blue. The first image shows the molecules with radii corresponding to the solvent accessible surface, i.e. $\alpha = 0$ (A). For larger $\alpha$-values more and more of the space between the solvent accessible surfaces gets covered C-D.}
\label{fig:alpha_filtration}
\end{figure}

The calculation of morphometric measures already relies on weighted Alpha complexes \cite{three_d_alpha_shapes, weighted_alpha_shapes, Koehl2023, alphamol}, which naturally provide a multi-scale representation of shape known as a filtration (see Fig. \ref{fig:alpha_filtration}). This Alpha-filtration is constructed on the union of atom centers of the simulated molecules. Persistent homology is the tool for quantifying topological features within such a filtration. 

We define a topological potential, $\mathcal{T}$, as a weighted sum of total persistence measures:
\begin{equation}
\label{topo_energy}
\mathcal{T} = \lambda_0 P_0 + \lambda_1 P_1 + \lambda_2 P_2,
\end{equation} 
where $P_i = \sum\limits_{(b,d)\in Dgm_i}(d-b)$ is the total persistence of all $i$-dimensional features (birth-death pairs) in the system's persistence diagram $Dgm_i$ and the $\lambda_i$ are scalar weights. 

The potential captures some shape information over arbitrary ranges suggesting a synergistic strategy in which the two potentials play complementary roles. The long-range topological term, $\mathcal{T}$, can create a guiding funnel within the energy landscape, pulling subunits into the correct general vicinity and orientation. The short-range geometric potential, $F_{sol}^*$, then acts as a refiner, selecting the precise low-energy binding interface. In this work, we test this principle by simulating dimer and trimer assembly of various systems governed by a combined potential: 
\begin{equation}
    \label{eq:interpolated}
    E_{comb} = \mu F_{sol}^* + (1-\mu) \mathcal{T}.
\end{equation}
We show that this approach, which combines local physical accuracy with global topological guidance, enhances the purely geometric model, increasing the simulation success rate for the TMV and SARS-CoV-2 ORF9b proteins by more than an order of magnitude and further making the successful assembly of a hepatitis B core protein and an extracellular fragment of the human CD40 ligand possible on the simulated timescales. This result validates the use of topology not just as a post-hoc analytical tool, but as an active biasing potential to solve kinetic challenges in molecular simulation.

\section{Results}
We begin discussing how $F^*_{sol}$ and $\mathcal{T}$ are correlated and analyze how persistence measures develop over the course of simulated assemblies driven only by $F^*_{sol}$. We do this to motivate the use of topological measures in our simulations using the inherent relationships between geometric and topological measures. We then investigate the minimizers of $\mathcal{T}$ for different combinations of $(\lambda_0, \lambda_1,\lambda_2)$. Finally, we compare different mixtures of $F^*_{sol}$ and $\mathcal{T}$ and how the inclusion of the topological potential improves the simulation results.

\subsection{Geometry-Topology Correlations}
\begin{figure*}[ht!]
\centering
\includegraphics[width=16.5cm]{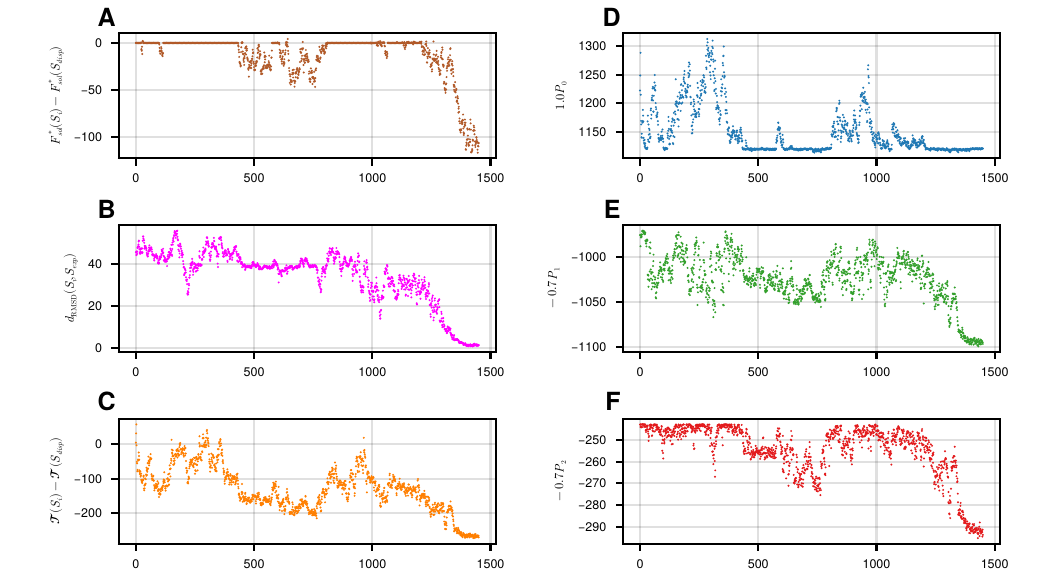}
\caption{Comparing values of a sequence of states obtained via a simulation driven by the geometric potential $F^*_{sol}$. The geometric potential itself (A). A goodness of fit measure with respect to the correct assembly of two TMV proteins (B). The weighted sum $\mathcal{T} = 0.5 P_0 -0.35 P_1 -0.35 P_2$ (C) The total persistence in dimension zero (D), dimension one (E) and dimension two (F).}
\label{fig:1_measure_paths_long}
\end{figure*}

We compare the topological and geometric potential, calculating $\mathcal{T}$ for a sequence of configurations of two TMV capsid protein subunits, obtained via a \textit{successful} simulation run driven by the geometric potential alone. Here successful means that the two subunits are assembled in a configuration similar to one observed experimentally, as the simulation reaches the energy minimum. Formally let $S_{ca}$ be an experimentally determined configuration of two correctly assembled protein subunits. Let $S$ be a configuration obtained via a simulation. We say that $S$ and $S_{ca}$ are similar if $d_{\text{RMSD}}(S, S_{ca}) < 2.5\text{\AA}$. See the Materials and methods section for an explanation of the distance measure $d_{\text{RMSD}}$. A visualization of the correct assembly of two protein subunits is shown in Figure \ref{fig:distance_phase} A. Each protein subunit is modeled as a rigid body and its geometry, i.e. the position of its atom centers, is based on the protein data base entry $\operatorname{6R7M}$ \cite{6R7M}. This entry also specifies how proteins are positioned in relation to one another in the experimentally observed helical assembly. For atomic radii we take values that subsume the hydrogen atoms into the atoms they are bound to \cite{protor}. The simulation is driven by a Random Walk Metropolis algorithm, see Materials and methods.

\subsubsection{Calculating total persistence for a successful simulation run}
Figure \ref{fig:1_measure_paths_long} A shows how $F_{sol}^*$, Equation \ref{fsol_star}, changes over the course of around $1500$ steps of a simulated assembly pathway. Here, a step of the assembly pathway corresponds to an accepted proposal of the Random Walk Metropolis algorithm, corresponding to a change in configuration space.
The signal is flat for large stretches of the simulation because the geometric potential is short-ranged and only exhibits changes in value when the solvent accessible surfaces of the simulated proteins overlap. There are several small valleys in the energy landscape, which correspond to configurations where the subunits are close. They detach again several times before finding a sequence of moves leading to a steep decline in energy and a local minimum, corresponding to configurations that are close to experimentally observed assembly. Figure \ref{fig:1_measure_paths_long} B shows the goodness-of-fit measure $d_{\text{RMSD}}$ that tracks how close to correct assembly the current configuration is. We can see that $d_{\text{RMSD}}$ is more stable in the regions where the subunits overlap and oscillates in the regions where $F_{sol}^*$ is flat. Around step $i = 1000$ there is a drop in $d_{\text{RMSD}}$ corresponding to the protein subunits being correctly aligned but not yet in contact. As $F_{sol}^*$ reaches its minimum, so does $d_{\text{RMSD}}$, which fluctuates around a low value indicating proximity to the correct assembly. 

Figure \ref{fig:1_measure_paths_long} D-F shows the weighted total persistence measures for dimensions zero, one and two. The chosen weights are $\lambda_0 = 1.0, \lambda_1 = -0.7, \lambda_2 = -0.7$. These weights correspond to the best performance increase achieved by mixing the geometric and topological potentials; see Section \ref{sec:guide}. The sum of all three, i.e. $\mathcal{T}_{(1.0, -0.7,-0.7)}$, is shown in Figure \ref{fig:1_measure_paths_long} C. The total persistence of the zeroth homology group is plotted in blue and is minimal as soon as the proteins are close together because then the zero-dimensional hole between the two rigid bodies is closed. It is interesting to note that $P_0$ is flat where $F_{sol}^*$ is not and vice versa. In addition, it seems that $P_2$ and $F_{sol}^*$ are correlated and that $d_{\text{RMSD}}$ and $\mathcal{T}_{(1.0, -0.7,-0.7)}$ share similar behavior. 

To quantify these observations, we calculate the Pearson coefficient \cite{pearson_coefficient} $r$ of the different functions. It is a measurement of the linear correlation between two variables $X = \{x_1, \dots, x_n\}$ and $Y= \{y_1, \dots, y_n\}$ and is given by \[
r = \frac{\sum\limits_{i=1}^n(x_i - \overline{x})\sum\limits_{i=1}^n(y_i - \overline{y})}{\sqrt{\sum\limits_{i=1}^n(x_i - \overline{x})^2}\sqrt{\sum\limits_{i=1}^n(y_i - \overline{y})^2}},
\]
where $\overline{x}$ and $\overline{y}$ are the averages of $X$ and $Y$. A value of $r = 1$ means perfect positive linear correlation, that is, if one variable increases, the other one increases proportionally. A value of $r=0$ means that there is no linear correlation, and a value of $r=-1$ means perfect negative linear correlation.
We list the Pearson coefficients of different measures of sequences of configurations of two TMV capsid proteins, in Table \ref{tab:pearson}. In the table, $\mathcal{T}$ refers to $\mathcal{T}_{(1.0, -0.7,-0.7)}$.

\begin{table}[ht!]
\centering
\caption{Linear correlations between different measures.}
\begin{tabular}{l|cccccc}
\toprule
 & $F_{sol}^*$ & $d_{\text{RMSD}}$ & $\mathcal{T}$ & $P_0$ & $P_1$ & $P_2$ \\
\midrule
$F_{sol}^*$ & $1.0$ & $0.71$ & $0.86$ & $0.38$ & $-0.8$ & $-0.91$ \\
$d_{\text{RMSD}}$ & $0.71$ & $1.0$ & $0.76$ & $0.44$ & $-0.7$ & $-0.75$ \\
$\mathcal{T}$ & $0.86$ & $0.76$ & $1.0$ & $0.42$ & $-0.98$ & $-0.95$ \\
$P_0$ & $0.38$ & $0.44$ & $0.42$ & $1.0$ & $-0.26$ & $-0.46$ \\
$P_1$ & $-0.8$ & $-0.7$ & $-0.98$ & $-0.26$ & $1.0$ & $0.87$ \\
$P_2$ & $-0.91$ & $-0.75$ & $-0.95$ & $-0.46$ & $0.87$ & $1.0$ \\
\bottomrule
\end{tabular}
\label{tab:pearson}
\end{table}

Linear correlations that stand out in particular are the high correlations of $\mathcal{T}_{(1.0, -0.7,-0.7)}$ to both $d_{\text{RMSD}}$ and $F_{sol}^*$, which are higher than the correlation between $d_{\text{RMSD}}$ and $F_{sol}^*$ themselves. The correlation of $d_{\text{RMSD}}$ to $\mathcal{T}_{(1.0, -0.7,-0.7)}$ is higher (in absolute terms) than to each individual persistence measure. Furthermore, there is a high negative correlation between $F_{sol}^*$ and $P_2$ of $-0.91$. The measures highly correlated in Table \ref{tab:pearson} match the plots in Figure \ref{fig:1_measure_paths_long} that exhibit similar trends. 

A final key observation is that $\mathcal{T}_{(1.0, -0.7,-0.7)}$ and the corresponding weighted persistence measures are at or close to their respective minima as the simulation reaches its assembled state. This in combination with the relatively high linear correlation between $d_{\text{RMSD}}$ and $\mathcal{T}_{(1.0, -0.7,-0.7)}$ motivates further analysis of the topological potential.

\subsection{Optimizing Persistence}

\begin{figure*}[ht!]
\centering
\includegraphics[width=16.5cm]{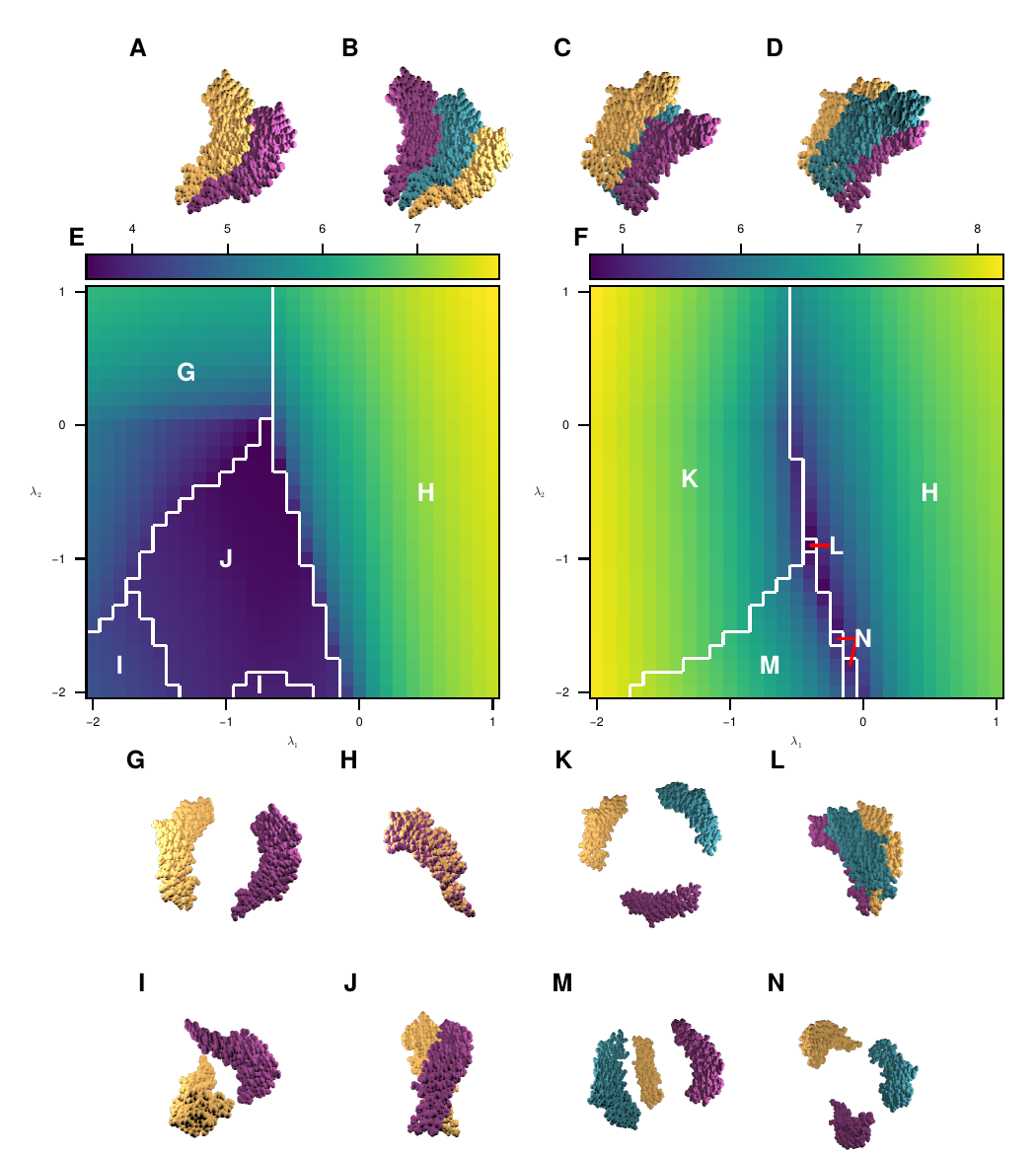}
\caption{Distances of correctly assembled states to the minimizer found for $\mathcal{T}$ in the parameter space as defined above. Render for two correctly assembled protein sub units (A), distance heatmap of correct assembly to minimizers (E), renders of the three different states for three sub units that we consider correctly assembled (B-D). Distance heatmap for the lowest distance between any of the three correctly assembled states for three subunits and the configurations of three subunits found to minimize $\mathcal{T}$ (F). Groups of topological potential minimizers (G-N) corresponding to the marked regions of the phase diagram. The topological potential minimizers closest to correct assembly are J and L.}
\label{fig:distance_phase}
\end{figure*}

We investigate the self-assembly of two and three subunits of the TMV protein under minimization of topological potential $\mathcal{T}$. Inferring promising weights for $\mathcal{T}$ from the correlation of $F_{sol}$ and $d_{\text{RMSD}}$ with the total persistence measures given in Table \ref{tab:pearson}, we analyze the parameter space given by $\lambda_0 = 1.0$, $\lambda_1 \in [-2.0, 1.0]$ and $\lambda_2 \in [-2.0, 1.0]$. We run both Random Walk Metropolis and simulated annealing simulations in a grid with step size $0.1$ in the parameter space. Let $\mathcal{S}$ be the set of all minimal configurations obtained during one of the simulations. For each point in the parameter space, we then check which configuration in $\mathcal{S}$ has the lowest topological potential. 
We then compare the topological potential of each minimizer to that of a correctly assembled state. 
We calculate \[
d_z = \min\limits_{S\in \mathcal{S}_{exp}} \operatorname{log}(\mathcal{T}_z(S) - \mathcal{T}_z(S_{z})).
\]
Here, $z = (\lambda_0, \lambda_1, \lambda_2)$ refers to a point in the parameter space and $S_z$ to the topological potential minimizer at $z$. $\mathcal{S}_{exp}$ is the set of assemblies observed experimentally. In the case of two subunits, $\mathcal{S}_{exp}$ contains the single configuration shown in Figure \ref{fig:distance_phase} A. In the case of three subunits, $\mathcal{S}_{exp}$ contains the configurations shown in Figure \ref{fig:distance_phase} B-D.

The resulting differences in topological potential $d_z$ for two and three subunits in the parameter space are plotted in Figures \ref{fig:distance_phase} E and F. Dark colors correspond to the region in which the differences between ground truth and found minimizer are small. The bounded regions in the plots correspond to different groups of topological potential minimizers that we grouped by distance measurement $d_{\text{RMSD}}$ and then manually aggregated further.

\subsubsection{Phases for two subunits}
In the case of two subunits, we can distinguish four large regions. One in which the subunits overlap almost perfectly is a large region to the right of the phase diagram, corresponding to configurations as shown in subfigure H. The second group are configurations in which the protein subunits do not touch. They appear in the upper left and lower left corners of the phase diagram and in the center bottom. We note that on the top left the subunits appear to form a circle between them, while on the bottom they appear to form a ball. This is not surprising considering that the one-dimensional persistence features contribute relatively more to the top-left, while the two-dimensional features, to the bottom. 

The final region lies roughly in the center of the phase diagram and contains configurations that touch and overlap slightly. None of the minimizing configurations is correctly assembled, although individual simulations did converge to a state close to correct assembly, which implies that it is a local minimizer. 

\subsubsection{Phases for three subunits}
The same images for three subunits are shown in Figure \ref{fig:distance_phase} K-N. Here, the region of non-touching subunits is much larger and roughly falls into three categories. Arranged such that a circle forms between the subunits as in subfigure K; a ball, as in subfigure M; and finally a spiral configuration, as in subfigure N. 

On the right side of the phase diagram there is the region with overlapping subunits, which is a bit smaller compared to the phase diagram of two subunits. This is caused by the possibility of more persistence features in dimensions one and two, counteracting the pull of minimizing zero-dimensional persistence. We have again marked this region with H, but note that here, three subunits overlap instead of two.

We again have a region in which the subunits touch but do not overlap completely, although it has almost disappeared. Interestingly, for this configuration two of the subunits are assembled in a way that corresponds to the correct assembly of two subunits, while the third one lies on top, with some overlap.

Considering the difference $d_z$ between the correct assembly and the topological potential minimizers, we see that for two subunits the region J, where the subunits touch but do not overlap corresponds to a dark area, that is, a small difference between the ground truth and the topological potential minimizer. For three subunits, this region of small $d_z$ values lies exactly at the boundary between the regions of minimizing configurations that do and do not overlap. The single point in the parameter space at $\lambda_1 = -0.4$ and $\lambda_2 = -0.9$ where the minimizer of three subunits touches but does not overlap is also the point in the parameter space with the lowest distance between ground truth and the topological potential minimizer.

\subsection{Combining geometry and topology}
\label{sec:guide}

We will now compare Random Walk Metropolis simulations in which a combination of geometric and topological potential is minimized. We recall the combined potential, Eq.\ \ref{eq:interpolated}:
\begin{equation*}
    E_{comb} =  \mu F_{sol}^* + (1-\mu) \mathcal{T},
\end{equation*}
where $\mu \in [0,1]$ is the interpolation between geometric and topological potential.

The measure in which we are interested is the success rate. Recall that a configuration $S$ is correctly assembled if $d_{\text{RMSD}}(S, S_{ca}) < 2.5 \text{\AA}$, for $S_{ca}$ denoting the experimentally observed assembly. Let $\mathcal{S}$ be a set of minimal configurations of a set of simulations, and let \[\mathcal{S}^* = \{S \in \mathcal{S} \mid d_{\text{RMSD}}(S, S_{ca}) < 2.5 \text{\AA}\}\] be the set of successful simulations. We define the ratio $\frac{\mid \mathcal{S}^* \mid}{\mid \mathcal{S} \mid}$ as the success rate of $\mathcal{S}$.

All simulations are run with the perturbation parameter $\sigma_t = 1.25$, which is the standard deviation for the normal distribution from which the translation perturbations are drawn; see the appendix. Figure \ref{fig:skewer} A, shows the success rates of simulations driven by $F_{sol}^*$ alone, depending on temperature $T$ and rotational perturbation parameter $\sigma_R$. Taking the value with the highest aggregated success rate in the temperature range, we set $\sigma_R = 0.2$ for all simulations driven by different mixtures $E_{comb}$.

\subsubsection{Weight-selection for the topological potential}
We select different combinations for triples of $(\lambda_0, \lambda_1, \lambda_2)$ defining $\mathcal{T}$. We pick $(1.0, -0.4, -0.9)$, which corresponds to the point in the parameter space where the minimizer of $\mathcal{T}$ for three subunits has touching but non-overlapping protein subunits. We further select $(1.0, -0.7, -0.7)$, which lies in the region of low distance between the correct assembly and the topological potential minimizer for two subunits. We chose $(1.0, 0.0, 0.0)$ to see how much of the impact can be attributed to the topological potential simply pulling the proteins together. And finally we pick $(1.0, -1.0, 1.0)$ because
it relates the topological potential to the integrated Gaussian curvature as: 

\[
\mathcal{T}_{(1.0,-1.0, 1.0)} = \frac{1}{4\pi}\int_{\alpha_0}^{\alpha_n}X(\mathcal{B}(\alpha)) - 1 \; d\alpha.
\] 
Here $\alpha_0$ and $\alpha_n$ are the lowest and highest value for which topological changes occur in the filtration of weighted Alpha complexes and $\mathcal{B}(\alpha)$ is the corresponding union of balls. See the appendix for details.

\begin{figure}[ht!]
\centering
\includegraphics[width=8.7cm]{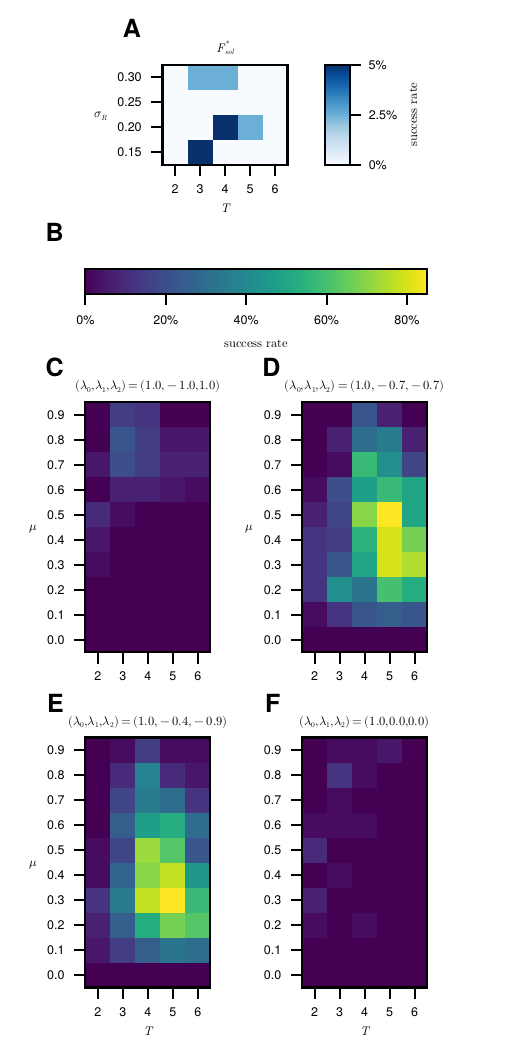}
\caption{Comparision of different simulation setups. Success rates for the pure morphometric approach, i.e. $(\lambda_0, \lambda_1, \lambda_2) = (0.0, 0.0, 0.0)$ depending on rotational perturbation parameter $\sigma_R$ and simulation temperature $T$ (A). A colorbar for the success rate for a mixture $E_{comb} = \mu F_{sol}^* + (1-\mu)\mathcal{T}$ (B). Heatmaps of the success rate depending on simulation temperature $T$ and weighting $\mu$ of topological potential $\mathcal{T}$, for parameters $(\lambda_0, \lambda_1, \lambda_2) = (1.0, -1.0, 1.0)$ (C), $(\lambda_0, \lambda_1, \lambda_2) = (1.0, -0.4, -0.9)$ (D), $(\lambda_0, \lambda_1, \lambda_2) = (1.0, -0.7, -0.7)$ (E) and $(\lambda_0, \lambda_1, \lambda_2) = (1.0, 0.0, 0.0)$ (F). }
\label{fig:skewer}
\end{figure}

\subsubsection{Scanning the parameter space for TMV dimerization}
Figure \ref{fig:skewer} C-F shows heat maps for the choices of $(\lambda_0, \lambda_1, \lambda_2)$ mentioned above, where we alter the simulation temperature $T =3,4,5,6$ and the factor $\mu = 0.1, 0.2, \dots, 0.9$ that interpolates between $F_{sol}^*$ and $\mathcal{T}$ to calculate the total energy as $E_{comb} = \mu F_{sol}^* + (1-\mu)\mathcal{T}$. For each point in the heatmap, we run $40$ simulations, where each simulation is run for $50000$ iterations. This corresponded to roughly $12$ hour long simulations on a single core of a high performance cluster. 

For the pure morphometric approach, the best success rate we achieve is $5\%$. The highest success rate among mixtures of both the morphometric approach and the topological potential is $85\%$ for $\mathcal{T}_{(1.0,-0.7,-0.7)}$ (Figure \ref{fig:skewer} D) at a simulation temperature of $5.0$ and a weighting $\mu = 0.5$, as well as for $\mathcal{T}_{(1.0,-0.4,-0.9)}$ (Figure \ref{fig:skewer} E) at $T=5.0$ and $\mu = 0.3$. This corresponds to a $16$-fold increase in the success rate. 

The highest success rate for $\mathcal{T}_{(1.0,-1.0,1.0)}$ (Figure \ref{fig:skewer} C) is equal to $22.5\%$. It is found for $T = 3.0$ and $\mu = 0.8$. Although substantially lower than for the other variants, it is still a $4$-fold increase over the pure morphometric approach. 

The combination of $F_{sol}^*$ with $\mathcal{T}_{(1.0,0.0, 0.0)}$ (Figure \ref{fig:skewer} F) achieves its best success rate of $12.5\%$ at $T = 3.0$ and $\mu = 0.8$. From this we infer that the improvements in success rate are not obtained only because zero-dimensional persistence is pulling the subunits together.

\begin{figure*}[ht!]
\centering
\includegraphics[width=16.5cm]{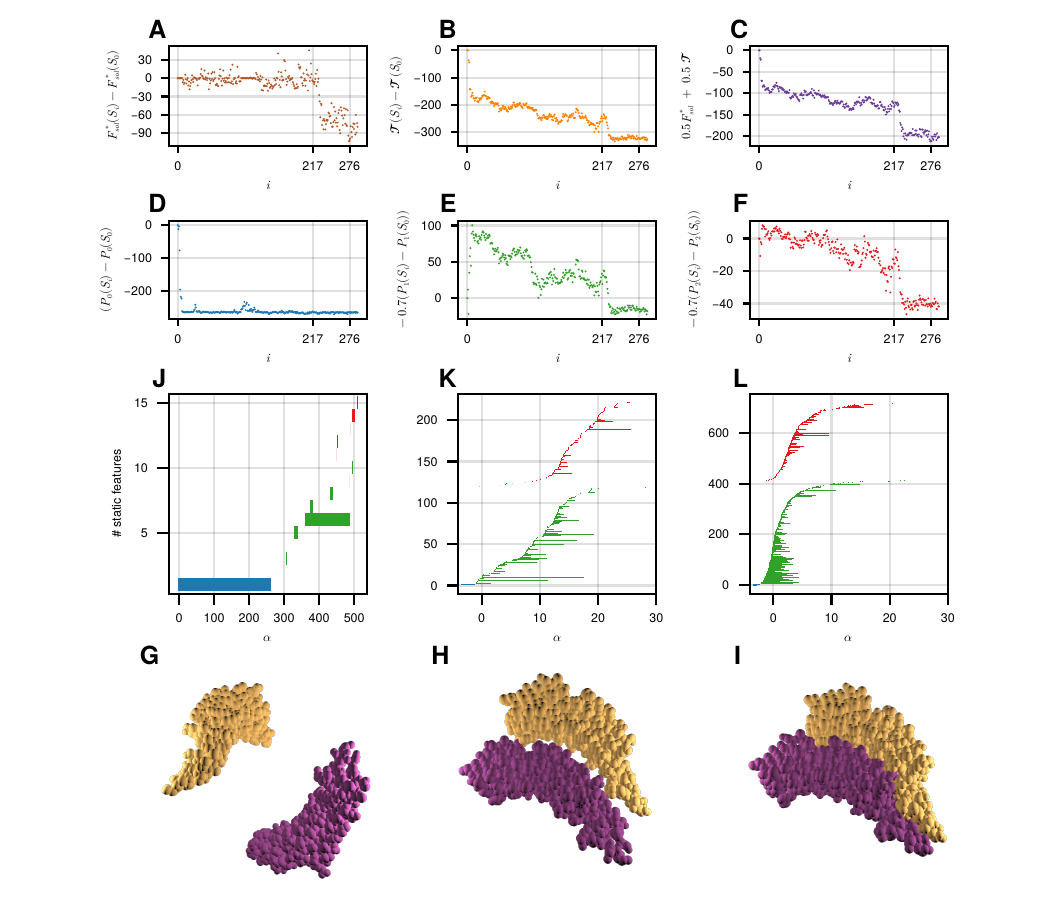}
\caption{Analysis of the final steps of a simulation driven by $E_{comb} = \frac{5}{10}F_{sol}^* + \frac{5}{10}\mathcal{T}_{(1.0, -0.7,-0.7)}$. 
All energy plots (A-F) show values relative to step $0$. We see $F_{sol}^*$ as given by Equation \ref{fsol_star} (A), $\mathcal{T}_{(1.0, -0.7,-0.7)}$ (B), $E_{comb}$ (C),
and the individual persistence measures (D-F). We show persistence bar plots of steps $0$, $217$ and $276$ with zero-dimensional features in blue, 
one-dimensional features in green and two-dimensional features in red (J-L). The corresponding protein configurations are shown in (G-I).}
\label{fig:persistence_in_ma_closeup}
\end{figure*}

\subsubsection{Analysis of a successful simulation of the combined potential}
Figure \ref{fig:persistence_in_ma_closeup} illustrates a sample simulated assembly pathway of two subunits with a mixture at $\mu = 0.5$ and $T=2.0$ for $\mathcal{T} =  P_0 - 0.7 P_1 - 0.7 P_2$. Again, steps of the assembly pathway refer to accepted proposals of the Random Walk Metropolis algorithm, corresponding to a change in configuration space. The simulation reaches its minimum energy, a correctly assembled configuration, after $276$ steps. The figures in subfigure A-C show $F_{sol}^*$, $\mathcal{T}$ and their mixture relative to the initial (dispersed) state. The D-E subfigures show the individual persistence measures. The D-E subfigures show the persistence barplots for three selected configurations. Namely, the initial state ($i = 0$), a state in which the subunits are correctly aligned but have a gap between them ($i = 217$) and finally the minimal energy state ($i = 276$), in which the subunits are correctly assembled. Renders of the corresponding configurations are shown in G-I. 

Note that $F_{sol}^*$ fluctuates around its initial value until step $217$, while the topological potential steadily declines. In the beginning most of the decline is driven by the disappearing zero-dimensional hole, but after about $20$ iterations the total persistence for dimension zero remains roughly constant, while the total persistence values for $P_1$ and $P_2$ decline. Interestingly, $P_1$ is almost as low as in the beginning as in the end, caused by a single very persistent feature. As the proteins get pulled together its value increases and then steadily declines. The same can be observed for $P_2$, although much less pronounced. This matches the observation from the phase diagram in Figure \ref{fig:distance_phase}, where configurations similar to the initial configuration are minimizers in parts of the phase diagram adjacent to the point in the parameter space we chose. After the initial period, where the proteins are pulled together, the decrease in topological potential is mainly driven by $P_1$ and supported by $P_2$. 

We refer to the sequence of steps before the simulation attains its minimum, but after the solvent accessible surfaces begin to overlap as the \textit{final phase}.

In the final phase, both the topological and geometric potentials rapidly decline, while the final gap between the protein subunits is closed. Considering just the topological potential, we observe that the final phase accounts for one third of the total decline, whereas the other two thirds occur during topologically driven alignment of the proteins. Looking at the persistence diagrams in subfigures J-L, we observe that more persistence features appear as the subunits get closer to the correct assembly and that the persistence features are born at lower $\alpha$ values and persist for a shorter time period. 

\subsubsection{Analyzing the impact of the topological potential in different simulation phases}
To quantify the impact of the topological potential at different stages of the assembly process, we compare the performance of $F_{sol}^*$ to $E_{comb} = \frac{1}{2}F_{sol}^* + \frac{1}{2}\mathcal{T}_{(1.0, -0.7, -0.7)}$ in a set of simulations that are initialized as the first state in the final phase of successful simulation runs. We take $20$ such states for both potentials, which are closest in $d_{RMSD}$ to the correct assembly. We then start ten simulations from each of the $40$ states for both potentials. Each simulation proposes $10000$ steps. The following Table \ref{tab:topo_power} shows the success rates for different combinations of initial states ($F_{sol}^*$ touch and $E_{comb}$ touch) and the energies driving the assembly ($F_{sol}^*$ finish and $E_{comb}$ finish). We compare the values of the combined potential at two different temperature values to verify the impact of potential detachment and realignment of the subunits.

\begin{table}
\centering
\caption{Comparing approximate success rates of $F_{sol}^*$ and $E_{comb}$ on first states of final phases of successful simulations for both potentials.}
\begin{tabular}{l|cc}
\toprule
  & $F_{sol}^*$ touch & $E_{comb}$ touch \\
\midrule
$F_{sol}^*$ finish, $T=3.0$ & $1\%$ & $6\%$ \\
$E_{comb}$ finish, $T=3.0$ & $25\%$ & $63\%$ \\
$E_{comb}$ finish, $T=5.0$ & $36\%$ & $54\%$ \\
\bottomrule
\end{tabular}
\label{tab:topo_power}
\end{table}

The key takeaway from the table is that the topological potential is useful in \textit{both} stages of the assembly process. That is, during long-distance alignment and short-range docking. $F_{sol}^*$ simulations initialized on states taken from successful $F_{sol}^*$ simulations find the correct assembly again in $1\%$ of the cases. $F_{sol}^*$ simulations initialized on states taken from successful $E_{comb}$ simulations correctly assemble in $6\%$ of the cases. This implies that alignment by $E_{comb}$ is better than alignment by $F_{sol}^*$. 

We simulate $E_{comb}$ for both $T=3.0$ and $T=5.0$ and see that in both cases simulations are more likely to succeed on the set of states that were aligned by $E_{comb}$. Interestingly, the lower temperature performs better in the $E_{comb}$ aligned states. We assume that this is caused by the selected states being comparatively good and that a higher temperature yields a higher chance to move away from them than a lower temperature. What is a drawback for well-aligned states is a feature for the low-quality alignments that appear during pure $F_{sol}^*$ simulations. Here, the higher temperature performs better, because moving away from the initialized alignment and realignment driven by $E_{comb}$ is favorable.

Overall, we conclude that the topological potential is a useful long-range augmentation of the morphometric approach because it draws the proteins together in a way that assists assembly and because it further supports assembly at close range.

\subsubsection{Improvements for other systems of protein assembly}
\begin{figure*}[ht!]
\centering
\includegraphics[width=16.5cm]{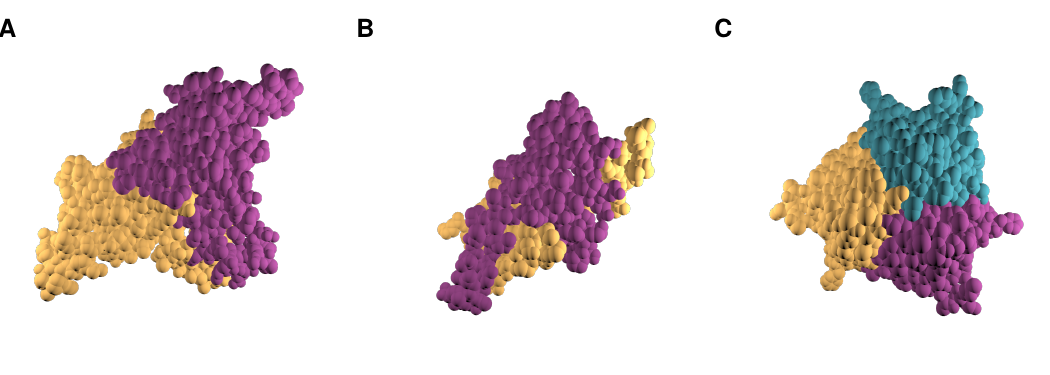}
\caption{Different simulated protein assemblies. Hepatitis B virus core protein (A), SARS-CoV-2 ORF9b (B) and fragment of the Human CD40 Ligand (C). Subunit geometry is given by the A-chains as specified in protein data base entries $\text{4BMG}$, $\text{6Z4U}$ and $\text{1ALY}$ respectively.}
\label{fig:others}
\end{figure*}

To show that this approach can be generalized, we apply it to simulate dimerization of the human hepatitis B virus core protein \cite{4BMG}, dimerization of an accessory protein of the SARS-CoV-2 virus that hinders the immune response \cite{6Z4U}, and trimerization of an extracellular fragment of the human CD40 ligand, relevant to T cell function \cite{1aly}. All of these structures are assemblies of homodimers and homotrimers, respectively, which means that they are composed of two or three identical protein chains. Figure \ref{fig:others} shows the renderings of the final assembled states found via simulations driven by 
\[E_{comb} = \frac{3}{10} F_{sol}^* + \frac{7}{10} \mathcal{T}_{(1.0, -0.4,-0.9)}.\]

We ran simulations for the pure morphometric approach and for combinations of $F_{sol}^*$ with $\mathcal{T}_{(1.0, -0.4,-0.9)}$ at $\mu = 0.3$ and $\mathcal{T}_{(1.0, -1.0,1.0)}$ at $\mu = 0.8$, corresponding to the parameters with the best performance, for the respective $\lambda_i$, observed in Figure \ref{fig:skewer}. We ran $40$ simulations each at simulation temperatures $T \in \{2.0, 3.0, 4.0, 5.0\}$ for two subunits and at $T = 5.0$ for three subunits.

Each simulation runs $50000$ and $60000$ iterations for two and three subunits respectively. Table \ref{tab:generalization_results} shows the highest success rate found for any temperature and the potentials described above. We include the previously discussed values of TMV dimerization for comparison. 

\begin{table}[ht!]
\centering
\caption{The success rate of simulations driven by the combined geometric-topological potential ($E_{comb}$) is compared against the purely geometric potential ($F_{sol}^*$) across four diverse protein systems. }
\begin{tabular}{llcccc}
\toprule
Protein System & Atoms & Success Rate & Success Rate & Success Rate \\
 & per subunit  & {$F_{sol}^*$} & {$\frac{8}{10} F_{sol}^* + \frac{2}{10} \mathcal{T}_{(1.0, -1.0,1.0)}$} & {$\frac{3}{10} F_{sol}^* + \frac{7}{10} \mathcal{T}_{(1.0, -0.4,-0.9)}$}  \\
\midrule
Tobacco mosaic virus & $1206$ & $5\%$ & $22.5\%$ & $85.0\%$ \\
Hepatitis B core &  $1134$ & $0\%$ & $12.5\%$ & $77.5\%$ \\
SARS-CoV-2 ORF9b & $639$ & $2.5\%$ & $7.5\%$ & $42.5\%$ \\
CD40 ligand fragment& $1113$ & $0\%$ & $0\%$ & $7.5\%$ \\

\bottomrule
\end{tabular}
\label{tab:generalization_results}
\end{table}

The key takeaways from the table are that the combined potentials perform better than the morphometric approach in all cases. In particular, for dimerization of the hepatitis B core and trimer assembly of the human CD40 ligand fragment, we observe an increase in the success rate from $0\%$ for the pure morphometric approach to $77.5\%$ and $7.5\%$ respectively, for the potential given by $\frac{3}{10} F_{sol}^* + \frac{7}{10} \mathcal{T}_{(1.0, -0.4,-0.9)}$. The success rate of dimerization of the SARS-CoV2-ORF9b protein jumps from $2.5\%$ to $42.5\%$ for the same potential. Judging by the values for the pure morphometric approach, the tobacco mosaic virus subunits appear easier to assemble, probably due to the high geometric complementarity compared to the hepatitis B core protein,
while simultaneously not being as intricately interlocked as the SARS-CoV2-ORF9b protein.

It is notable that the selection of parameters that yielded the best results for the tobacco mosaic virus also yields strong improvements for other systems. The fact that it performs better for the hepatitis B core than the SARS-CoV2-ORF9b protein is likely influenced by the greater difference in the size of the system, that is, the atoms per subunit, as shown in column 3 of Table \ref{tab:generalization_results}. We note that in all cases where we observed at least one simulation that obtained a minimum corresponding to the biologically correct assembly, this structure was also the configuration minimizing $E_{comb}$ compared to all other simulation results. As such the combined potential is thermodynamically sound in the sense that it correctly distinguishes the ground truth as the energetically most stable state. 

\section{Discussion}

Our study demonstrates that a topological potential, defined as the weighted sum of total persistence measures of proteins in a solvent, substantially improves the success rate of simulated protein assemblies. Acting as a long-range complement to the short-range morphometric approach to solvation-free energy, it enables assembly in systems that fail under the morphometric potential alone and increases success rates by more than an order of magnitude for other systems, reaching $85\%$ for TMV dimerization with optimized parameters.

We analyze the relationship between the topological potential
$\mathcal{T}$ and the morphometric solvation term with a soft-sphere constraint $F_{sol}^*$, in depth for TMV assembly. Parameters optimized for TMV translate well to other systems, particularly those with protein subunits of comparable atomic size, yielding high success rates without system-specific tuning.

While the topological potential does not yet have a physical interpretation, one specific parameterization, using the alternating sum of total persistence measures, can be mathematically linked to the Gaussian curvature term in the morphometric approach. This may be understood as capturing topology changes in an expanding molecular surface, potentially connected to solvation shell organization or other long-range, shape-dependent effects.

Because $\mathcal{T}$ depends only on atomic coordinates, it can be applied to any in silico self-assembly problem where these are known and can be combined with arbitrary potentials. Integrating such a shape-based, long-range term with physically grounded short-range models offers a computational strategy for overcoming rugged energy landscapes in molecular assembly simulations, with demonstrated applicability across diverse protein self-assembly systems and potential relevance to other biological assemblies and materials science.

\section*{Acknowledgments}
Funded by the Deutsche Forschungsgemeinschaft (DFG - German Research Foundation) - Project-ID 195170736 - TRR109. We acknowledge Patrice Koehl for the sharing and explanation of his program AlphaMol. We thank Gero Friesecke for discussions on virus self-assembly. We acknowledge Fulbright Germany for financing a fellowship at the Lawrence Berkeley National Laboratory for Ivan Spirandelli. This research used the computational cluster resource provided by the Center for Information Technology and Media Management at the University of Potsdam.
This research used the Lawrencium computational cluster resource provided by the IT Division at the Lawrence Berkeley National Laboratory.
This work was supported by the Director, Office of Science, Office of Advanced Scientific Computing Research, of the U.S. Department of Energy under Contract No. DE-AC02-05CH11231.

\bibliographystyle{my_unsrtnat}
\bibliography{references}

\newpage
\section*{Appendix}
\label{sec:mm}

\noindent
\subsection*{Persistent homology of Alpha filtrations}

The geometric and topological measures are derived from the weighted Alpha complex of the atomic centers. The Alpha complex is a subcomplex of the weighted Delaunay triangulation that provides a mathematically precise and computationally efficient representation of the union-of-balls model of a molecule \cite{three_d_alpha_shapes, weighted_alpha_shapes}. For a set of atomic centers $P = \{p_1, \dots, p_n\} \subset \mathbb{R}^d$ and a corresponding set of weights $W = \{w_1, \dots, w_n\} \subset \mathbb{R}$, the construction begins with the \textit{weighted Voronoi diagram}. The $i$\textit{-th weighted Voronoi cell} is defined as
\[
    V(p_i, w_i) := \{x \in \mathbb{R}^d \mid \pi_{p_i, w_i}(x) \leq \pi_{p_j, w_j}(x), \forall j \in \{1, \dots, n\}\},
\]
where $\pi_{p, w}(x) = ||x-p||^2 - w$ is the power distance. The collection of weighted Voronoi cells is the weighted Voronoi diagram denoted by $\mathcal{V}(P,W)$.

Under the assumption that no $d+2$ Voronoi cells intersect in a common point, i.e., the points in $P$ are in general position, the dual to the weighted Voronoid diagram is the \textit{weighted Delaunay triangulation}. 

Now let $W = \{w_1, \dots, w_n\} \subset \mathbb{R}$ be a set of weights assigned to the points in $P$, where each weight $w_i$ is the squared atomic radius ($w_i = r_i^2$). Let $B(p_i,\sqrt{w_i})$ be the ball centered at point $p_i$ with radius $r_i = \sqrt{w_i}$. We denote \[R(p_i, w_i) = V(p_i,w_i) \cap B(p_i,\sqrt{w_i}).\]
The \textit{weighted Alpha complex} is defined as \[
    \mathcal{A}(P,W) = \{\sigma \subseteq P \mid \bigcap\limits_{p_i \in \sigma} R(p_i, w_i) \neq \emptyset\},
\]
i.e. the dual of the weighted Voronoi cells intersected with the balls $B(p_i,\sqrt{w_i})$.
Inclusion-Exclusion formulas to compute the geometric measures $V$, $A$, $C$ and $X$, are based on this incidence information \cite{union_of_balls, vol_derivative, area_derivative, mean_derivative, gaussian_derivative}. These formulas are implemented in the AlphaMol program \cite{alphamol}. We also use the weighted Alpha complex to compute the overlap penalty $L$, by iterating over the edges whose vertices correspond to different proteins and summing up the overlap of the corresponding balls.

Let $\alpha \in \mathbb{R}$. We define $B(p_i, w_i, \alpha)$ as the ball centered at $p_i$ with radius $r_i = \sqrt{w_i + \alpha}$, where $B(p_i, w_i, \alpha) = \emptyset$ if $\alpha < -w_i$. 
We get that for points $x$ with equal power distance to balls $i$ and $j$: \[
    \pi_{p_i, r_i^2}(x) = ||x-p_i||^2 - (w_i + \alpha) = ||x-p_j||^2 - (w_j + \alpha)  = \pi_{p_j, r_j^2}(x),
\]
which is equivalent to $||x-p_i||^2 - w_i =  ||x-p_j||^2 - w_j$ as the $\alpha$ cancels out. 
Therefore, the weighted Voronoi diagram is the same for all values of $\alpha$. We write \[W_\alpha = \{w + \alpha \mid w \in W\}\]
and get a nested sequence of complexes $$\mathcal{A}(P, W_{\alpha_0}) \subset \dots \subset \mathcal{A}(P, W_{\alpha_m})$$
for critical values of alpha $\alpha_0 \leq \dots \leq \alpha_m$. The critical values are the lowest values for which a simplex in the weighted Delaunay Triangulation appears in the corresponding weighted Alpha complex. 

Given a filtration $F = \mathcal{A}_0 \subset \dots \subset \mathcal{A}_m$ and homomorphisms, induced by the inclusion maps, \newline \noindent$f_p^{i,j}: H_p(\mathcal{A}_i) \to H_p(\mathcal{A}_j)$, the images
\[
    H_p^{i,j} = \operatorname{im}(f_p^{i,j})
\]
are the $p$-dimensional \textit{persistent homology groups}. Their ranks \[
    \beta_p^{i,j} = \operatorname{rank}(H_p^{i,j}),
\]
are the $p$-dimensional \textit{persistent Betti numbers}. They count the $p$-dimensional cycles that are not boundaries.
Given a class $\gamma \in H_p(\mathcal{A}_i)$ we say $\gamma$ is \textit{born} at $\mathcal{A}_i$ if $\gamma \notin H_{p}^{i-1, i}$. We say it \textit{dies}
at $\mathcal{A}_j$ if $f_p^{i,j-1}(\gamma) \notin H_{p}^{i-1,j-1}$ and $f_p^{i,j}(\gamma) \in H_{p}^{i-1,j}$. We call the interval $(\alpha_i,\alpha_j)$ the \textit{birth-death interval} of $\gamma$ and $\alpha_j-\alpha_i$ its \textit{persistence}.
We refer to the collection of all $p$-dimensional birth-death intervals as the $p$-dimensional \textit{persistence diagram} and denote it by $Dgm_p$ \cite{Computational+Topology}. 

We use the Oineus software package \cite{oineus} to calculate persistent homology.  \newline

\noindent
\subsection*{Simulation methods and parameters}

We simulate configurations of protein subunits using variants of a Random Walk Metropolis \cite{metropolis}. Each configuration is represented by a state $x \in (\operatorname{SO}_3 \times \mathbb{R}^3)^m$, where $m$ is the number of simulated molecules. The geometries of the systems we study are taken from the protein data base entries with identifiers 6R7M \cite{6R7M}, 4BMG \cite{4BMG}, 6Z4U \cite{6Z4U} and 1ALY \cite{1aly}, for TMV, hepatitis B, SARS-CoV2 and the CD40 ligand respectively. For atomic radii we use the \textit{ProtOr} radii \cite{protor}, which account for hydrogen atoms by giving radii for atomic groups instead of individual atoms. 
Each simulation is run in a bounding box of size $120\text{\AA}$. These bounds are not checked for each individual atom but for each center of mass of the simulated molecules. 

Perturbation for a molecule consist of a translation and a rotation. The translational perturbation vector is generated by drawing three independent samples from a normal distribution $\mathcal{N}(0, \sigma_t^2)$ with $\sigma_t = 1.25\text{\AA}$, and this vector is added to the molecule's current position. The rotational perturbation is generated by first creating a three dimensional random vector $v$, whose components are also drawn independently from a normal distribution, $\mathcal{N}(0, \sigma_R^2)$, with $\sigma_R = 0.2$. This vector $v$ represents an element of the Lie algebra $\mathfrak{so}(3)$. It is then mapped to a rotation in $\mathrm{SO}(3)$ via the exponential map. This resulting small, random rotation is then composed with the molecule's existing orientation to yield the new candidate orientation.

Assuming the current state of the simulation is state $x^i$, we pick one of the simulated molecules at random and perturb it to generate a candidate state $x'$. This is also known as \textit{component-wise} or \textit{block-wise} Random Walk Metropolis \cite{cc_rwm}.

The probability of accepting $x'$ as the new state of the simulation is determined as:
\[
    \operatorname{min}\left(1, \operatorname{exp}\left(\frac{-\Delta E}{T}\right)\right),
\]
where $\Delta E = E(x') - E (x^i)$ is the difference in energy of the states. If $x'$ is accepted, we set $x^{i+1} = x'$. Otherwise we set $x^{i+1} = x^{i}$. 
$T$ is a simulation temperature, which regulates the probability of accepting states with higher energy. The energy used here is $E_{comb}$ as in Equation \ref{eq:interpolated} for different parameters $\mu$ and $\lambda_i$. The prefactors of $F_{sol}^*$ are given by the White Bear prefactors \cite{white_bear} computed for packing fraction $\eta = 0.3665$ and solvent radius $r_s = 1.4\text{\AA}$, mimicking the properties of water.
For the simulated annealing (SA) \cite{simulated_annealing_1} simulations we change $T$ over the course of a simulation by letting it decay to zero according to a quadratic annealing schedule: 
\[
    T_k = T_0 \left(\frac{I-k}{I}\right)^2,
\]
where $I$ is the total number of iterations at which the temperature reaches zero and $k$ is the current step in the decay period. Once $T = 0.0$ is reached, we reheat the system several times to scan for nearby local minima.

In cases in which we did not know which initial temperature to start a simulation with, we run $12$ shorter RWM simulations and check whether the total acceptance rate is above or below some target acceptance rate. If it is below the target acceptance rate, we multiply the initial temperature for the next simulation by a factor of $1.5$, if it is above the target acceptance rate we multiply the initial temperature for the next simulation by a factor of $0.5$. After the $12$ simulations are finished we pick the initial temperature closest to our target acceptance rate. For RWM simulations we choose a target acceptance rate of $0.2$ and for SA simulations we choose a target acceptance rate of $0.8$.

The morphometric approach is short ranged and additive. As such it is not necessary to recompute its energetic contribution of the whole system in every step, but only the connected components of molecules that have changed. We determine the connected components in each step by constructing a graph that has each molecule as a node and an edge whenever the bounding spheres of two molecules overlap. Let $idx$ be the index of the molecule we moved for the proposal of simulation step $i+1$. Let $\mathcal{C}_i$ be the set of connected components from step $i$ and let $\mathcal{C}_{i+1}$ be the set of connected components from step $i+1$.
For each component $c \in \mathcal{C}_{i+1}$ there are three cases to consider.  
Case 1: The connected component $c$ is an element of $\mathcal{C}_i$ and $idx \notin c$. 
Then the energy of $c$ remains unchanged and we can just copy the energy value from the last iteration.
Case 2: The connected component $c$ is an element of $\mathcal{C}_i$ and $idx \in c$. 
Then the energy of $c$ changes and we have to recompute it. 
Case 3: The connected component $c$ is not an element of $\mathcal{C}_i$.
Then the energy of $c$ is changed and we need to recompute.
By passing the energy value of a single molecule to the algorithm we can further optimize this procedure and not recalculate the energetic contributions of components that consists of only one molecule.

To ensure a fair comparison between different topological potential setups, we defined a relative contribution measurement, $c_{\mathcal{T}}$. This measurement quantifies the energetic impact of the topological term relative to the geometric term during the last stage of the assembly process. We take the set of $800$ simulations, the data shown in Figure \ref{fig:skewer} A are based on. For every simulation $i$ we take two states $S^i_{min}$, the minimal energy state, and $S^i_{disp}$, the state immediately before the solvent-accessible surface areas of the protein subunits start to interact, before $S^i_{min}$ is reached. We then calculate an average of how much of the change in energy in the final phase in these simulations is contributed by $\mathcal{T}$ in relation to $F_{sol}^*$ as: \[
    c_{\mathcal{T}} = \frac{1}{800}\sum\limits_{i=1}^{800} \frac{\mathcal{T}(S^i_{min}) - \mathcal{T}(S^i_{disp})}{F_{sol}^*(S^i_{min}) - F_{sol}^*(S^i_{disp})}.
\] 
We find that $c_{\mathcal{T}_{(1.0, -0.7, -0.7)}} \approx 1.12$, $c_{\mathcal{T}_{(1.0, -0.4, -0.9)}} \approx 0.96$ and $c_{\mathcal{T}_{(1.0, -1.0, 1.0)}} \approx 0.61$, i.e. the values are reasonably similar for the different setups, ensuring comparability. Note that calculating $c_{\mathcal{T}_{(1.0, 0.0, 0.0)}}$ is not very informative because zero dimensional persistence has little impact in the final stage of a simulation. \newline

\subsection*{Relation between the topological potential and the integrated Gaussian curvature}

Consider a set of atom centers $P = \{p_1, \dots, p_n\}$ and van der Waals radii $R = \{r_1, \dots, r_n\}$. The set of weights $$W = \{(r_j + r_s)^2 \mid r_j \in R\}$$ for the solvent radius $r_s$ are the weights
of the weighted Alpha complex used to compute the morphometric measures. 
Now let $\mathcal{B}(\alpha) =\bigcup\limits_{j=1}^mB_j(\alpha)$ where $B_j(\alpha)$ is the ball centered at $p_j$ with radius $\sqrt{w_j + \alpha}$. If $w_j+\alpha<0$, we define $B_j(\alpha) = \emptyset$.
Because the Alpha complex $\mathcal{A}_\alpha$ is homotopy equivalent to the union of balls $\mathcal{B}(\alpha)$, their Euler characteristics are equal \cite{union_of_balls}. The sequence of values, $\alpha_0 < \dots < \alpha_k$ for which topological changes occur, when moving from $\mathcal{A}(\alpha_i)$ to $\mathcal{A}(\alpha_{i+1})$, are the filtration values of the Alpha filtration on which we calculate persistent homology. Here $$\alpha_0 = -\max\limits_{j\in 1,\dots,m} w_j.$$
The Euler characteristic $\chi$ of a union of balls $\mathcal{B}(\alpha)$ is related to its integrated Gaussian curvature $X$ (integrated over the boundary) by a factor of $4\pi$. That is: \[
    X(\mathcal{B}(\alpha)) = 4\pi\chi(\mathcal{B}(\alpha)).
\]
Then by the Euler-Poincaré formula \cite{Computational+Topology} we get: 
\[
    \sum\limits_{i=0}^2(-1)^i\sum\limits_{(b,d)\in Dgm_i}(d-b) = \int_{\alpha_0}^{\alpha_n}\chi(\mathcal{B}(\alpha)) - 1 \; d\alpha.
\] 
That is, the alternating sum of total persistence per dimension is equal to the integral of $\alpha_0$ to $\alpha_n$ over the Euler characteristic $\chi(\mathcal{B}(\alpha))$. We subtract $1$ in the integral because the connected component representing the whole molecule persists forever, and we do not count it in the alternating sum of the total persistence values. 

Therefore: 
\[
\mathcal{T}_{(1.0,-1.0, 1.0)} = \int_{\alpha_0}^{\alpha_n}\frac{1}{4\pi}X(\mathcal{B}(\alpha)) - 1 \; d\alpha.
\]

This choice of parameters can thus be seen as a range extension of one of the terms of the morphometric approach in a mathematically rigorous sense. 

We note that for $\alpha \leq 0$, we expect the integral to be roughly constant across different configurations, since the atom centers within a protein subunit are usually closer together than those between subunits.
For $\lambda_0^* = \lambda_0-1$, $ \lambda_1^*=\lambda_1 +1$ and $\lambda_2^* = \lambda_2-1$ we have that: 
\[
    \sum\limits_{i=0}^2 \lambda_i\sum\limits_{(b,d)\in Dgm_i}(d-b) = \sum\limits_{i=0}^2((-1)^i + \lambda_i^*)\sum\limits_{(b,d)\in Dgm_i}(d-b),
\]
meaning the topological potential $\mathcal{T}$, can always be decomposed into:
\[
\mathcal{T}_{(\lambda_0, \lambda_1, \lambda_2)} = \int_{\alpha_0}^{\alpha_n}\frac{1}{4\pi}X(\mathcal{B}(\alpha)) - 1 \; d\alpha + \mathcal{T}_{(\lambda_0^*, \lambda_1^*, \lambda_2^*)}
\]
Hence, while the morphometric approach has the integrated Gaussian curvature $X(\mathcal{B}(0))$ as a component, the topological potential contains the integral over the Euler characteristic of $\mathcal{B}(\alpha)$ over the range of $\alpha$ values in which topological changes occur in the Alpha filtration.
\newline

\subsection*{Ground truths from multiple A-chains}

The data provided in the protein data base files $\text{4BMG}$, $\text{6Z4U}$ and $\text{1ALY}$ contain geometric information for the protein assemblies discussed in Figure \ref{fig:others} and Table \ref{tab:topo_power}. The dimers or trimers assemblies of several proteins of the same type, but have subtle differences in the geometry of each chain, due to the non-rigid nature of the proteins or measurement imprecision. We take the A-chain of all entries as the geometric template in the simulations we run. The ground truth to which we compare simulation results is obtained by superimposing the A-chain onto the positions and orientations of the other chains of the dimer or trimer.
\newline

\subsection*{Simulation Evaluation}

Let $Q,U,V,W \subset \mathbb{R}^{3k}$ be rigid transformations of some underlying template $X_0 \subset \mathbb{R}^{3k}$ representing a molecule with $k$ atoms.

Now assume that $Q$ and $U$ are a pair of molecules in some configuration $\mathcal{X}$ and that $V$ and $W$ are a pair of molecules in some configuration $\mathcal{Y}$. Let $(R_{Q \to V}, t_{Q \to V})$ be the optimal rigid-body transformation that superimposes molecule $Q$ onto molecule $V$. This transformation is found by solving the minimization problem:
\[
    (R_{Q \to V}, t_{Q \to V}) = \argmin\limits_{R \in SO(3), t \in \mathbb{R}^3} \sum_{i=1}^k \left\| (R q_i + t) - v_i \right\|^2.
\]
This problem can be solved efficiently using the Kabsch algorithm \cite{kabsch_1978}.

Using this transformation, we define a pairwise RMSD, $d_{\text{pair}}$. This metric quantifies the structural deviation of the molecule, $U$, 
from its target, $W$, after the transformation $(R_{Q \to V}, t_{Q \to V})$ has been applied.
Because we assume all molecules are perfect rigid-body instances of the same template, the initial alignment of $Q$ onto $V$ results in zero error. 
The metric is therefore defined solely by the deviation of the second molecule:
\[
    d_{\text{pair}}(Q, U, V, W) = \sqrt{\frac{1}{k} \sum_{i=1}^k \left\| (R_{Q \to V} u_i + t_{Q \to V}) - w_i \right\|^2}.
\]
Note that the sum is now over the $k$ atoms of the second molecule only, and the normalization factor is $k$. 

To handle the ambiguity of local assignment within the pair, the final pairwise distance, $d_{\text{minpair}}$, 
is the minimum of the two possible alignments, $Q$ to $V$ or $Q$ to $W$:
\[
    d_{\text{minpair}}(Q, U, V, W) = \min\left( d_{\text{pair}}(Q, U, V, W), d_{\text{pair}}(Q, U, W, V) \right).
\]
Finally, the total distance $d_{\text{RMSD}}$ between two configurations, $\mathcal{X} = \{X_1, \dots, X_m\}$ and $\mathcal{Y} = \{Y_1, \dots, Y_m\}$, is the average of these pairwise distances, minimized over permutations $\pi \in S_m$:
\[
    d_{\text{RMSD}}(\mathcal{X},\mathcal{Y}) = \min_{\pi \in S_m} \frac{1}{\binom{m}{2}} \sum_{1 \le i < j \le m} d_{\text{minpair}}(X_i, X_j, Y_{\pi(i)}, Y_{\pi(j)}),
\]
where the sum goes over all unique pairs of molecules in the configuration.

Intuitively, $d_{\text{RMSD}}$ finds the optimal global matching of molecules. For each matched pair, 
it determines the best alignment based on one of the molecules and then measures the resulting RMSD of the entire pair. 
A $d_{\text{RMSD}}$ value of zero guarantees that two configurations are identical up to a global rigid transformation. 
\newline

\subsection*{Codebase}

The code we developed to generate the data presented in this study is available on github \cite{morphomol}.

\end{document}